\documentclass[showpacs,superscriptaddress,amsmath,twocolumn,floatfix,prl]{revtex4}

\usepackage{subfigure}
\usepackage{graphicx}
\usepackage{amssymb}

\begin{document}

\title{Giant vortices, vortex rings and reentrant behavior in type-1.5 superconductors}

\author{V. H. Dao}
\author{L. F. Chibotaru}
\affiliation{INPAC - Institute for Nanoscale Physics and Chemistry,\\
Katholieke Universiteit Leuven,  Celestijnenlaan 200F, B-3001 Leuven, Belgium }

\author{T. Nishio}
\affiliation{Department of Physics,
Tokyo University of Science, Kagurazaka, Shinjuku, Tokyo 162-8601, Japan }

\author{V. V. Moshchalkov}
\affiliation{INPAC - Institute for Nanoscale Physics and Chemistry,\\
Katholieke Universiteit Leuven,  Celestijnenlaan 200F, B-3001 Leuven, Belgium }
%\date{today}

\begin{abstract}
We predict that in a bulk type-1.5 superconductor the competing magnetic responses of the two components of the order parameter can result in a vortex interaction that generates group-stabilized giant vortices and unusual vortex rings in the absence of any extrinsic pinning or confinement mechanism. We also find within the Ginzburg-Landau theory a rich phase diagram with successions of behaviors like type-1$\rightarrow$type-1.5$\rightarrow$type-2$\rightarrow$type-1.5  as temperature decreases.   
\end{abstract}


\maketitle

Topological defects have been intensively investigated in a wide variety of physical systems in e.g. condensed matter (ferromagnets, superfluids, liquid crystals, etc.) \cite{mineev}, optics \cite{weiss}, particle physics or astrophysics \cite{rajantie}, which has given rise to fruitful exchanges of ideas between the different fields. In a superconductor the quantized line defect is a vortex (also called flux line) which carries a localized integer multiple of the magnetic flux quantum $\Phi_0=h/2e$. Close to the transition temperature a unique quantity, the Ginzburg-Landau (GL) parameter $\kappa$, determines superconductivity response to a magnetic field \cite{brandt}. It is defined as the ratio of the magnetic penetration depth $\lambda$ to the coherence length $\xi$ of the order parameter (OP) and it can be calculated from the microscopic parameters of the material within the BCS theory. In a type-1 superconductor, $\kappa<1/\sqrt{2}$, the interaction between vortices is purely attractive which results in their fusion into macroscopic normal domains in the intermediate state. On the contrary in a type-2 superconductor,  $\kappa>1/\sqrt{2}$, the interaction is purely repulsive so the vortices are stable and form a lattice of co-directed lines in the mixed state. At the critical value the GL theory (which is mathematically equivalent to the abelian Higgs model in particle physics) predicts that the vortex interaction is exactly zero \cite{bogomolnyi,jacobs}. However experiments with $\kappa\approx 1/\sqrt{2}$ observed complex patterns consisting of mixed-state domains with vortex-free Meissner domains \cite{krageloh,brandt}. This stems from a non-monotonic inter-vortex potential which is repulsive at short separation and attractive at a long range. Such interaction profile can be explained theoretically by taking into account different mechanisms such as low-temperature corrections \cite{brandt,lukyanchuk} or fluctuations and anisotropy in the vortex lattice \cite{blatter}. 

Recently Babaev and Speight \cite{babaev} predicted that in a superconductor with two uncoupled components the inter-vortex potential can be non-monotonic when one component of the OP is intrinsicly of type 1 while the other is of type 2. This idea has been recently supported by observations \cite{moshchalkov} of highly inhomogeneous vortex distributions in clean samples of two-band superconductor MgB$_2$ \cite{nagamatsu}. 
The estimates of MgB$_2$ parameters \cite{moshchalkov} indicate that this behavior, coined type-1.5 superconductivity, only occurs in high quality samples which explains why the material was classified before as type-2. Including the finite inter-band Josephson coupling (which is moderate in MgB$_2$), GL calculations \cite{moshchalkov} yielded non-monotonic vortex interaction potentials, also when one of the bands is inactive. A detailed study of this inter-band 'proximity effect' \cite{babaev2} has shown that a type-1.5 potential is then associated with  dissimilar variations of the OP components at finite distances from the vortex core, despite their long-range behaviors being identical.  Type-1.5 superconductivity could be more commonly observed than the secondary effects showing up in single-component superconductors with $\kappa= 1/\sqrt{2}$. Stimulated by the growing number of discovered multi-band superconductors, e.g. iron-based  superconductors \cite{kamihara,hsu}, we investigate in this article novel features of the type-1.5 regime which distinguish it from single-component superconductivity at the critical $\kappa$. We find, in particular, group-stabilized giant vortices and vortex rings composed of single/multi-quantum vortices as thermodynamically stable phases.

\emph{The two-band GL theory.\textemdash} In the context of two-band superconductivity with s-wave symmetry of the OP we consider the GL free-energy functional  \cite{dao} 
\begin{equation}
{\cal F}_{GL}= {\cal F}_1 + {\cal F}_2 - \gamma (\Psi_1^* \Psi_2 \!+\! \Psi_1 \Psi_2^* ) + \frac{1}{8\pi}(\nabla \times {\bf A})^2
\label{eq:FGL}
\end{equation}
where the contributions from each band
\begin{equation}
{\cal F}_n = \alpha_n |\Psi_n|^2 + \frac{\beta_n}{2} |\Psi_n|^4  + \frac{1}{2 m_n}|(-i\nabla + \frac{2 \pi}{\Phi_0} {\bf A})\Psi_n|^2 
\end{equation}
 are supplemented by a Josephson-type coupling and the magnetic energy. The Josephson interaction describes Cooper pair tunneling between the two bands. It is the simplest of the terms mixing the two components $\Psi_n$ of the OP and in MgB$_2$, for example, the other ones vanish in the clean limit~\cite{dao,gurevich}. Although $\alpha_1$ and $\alpha_2$ can change sign at different temperatures, any finite Josephson coupling $\gamma$ imposes that both $\Psi_n$ vanish at the same critical temperature $T_c$. Below $T_c$ in the homogeneous Meissner state ${\bf A}=0$ and $|\Psi_n|= u_n$, which are solutions of $u_n (\alpha_n + \beta_n u_n^2) - \gamma u_{3-n} =0$. The energy density is then 
\begin{equation}
 \epsilon_M= - (\beta_1 u_1^4 + \beta_2 u_2^4)/2.
\end{equation}

Superconductivity with two OP components is richer than with a single one because not only the ratio $u_2/u_1$ changes with temperature but also the two components can have different magnetic responses and spatial variations. Let us introduce the band parameters $\kappa_n$ and their average $\bar{\kappa}$ defined by
\begin{eqnarray}
\kappa_n & = & \Phi_0 m_n \sqrt{\beta_n}/(2 \pi)^{3/2} \\
\bar{\kappa}^{-1} & = & \sqrt{p_1} \kappa_1^{-1} + \sqrt{p_2} \kappa_2^{-1}
\end{eqnarray}
where the condensation-energy fraction 
\begin{equation}
p_n = \beta_n u_n^4/2|\epsilon_M|.
\end{equation} 
If the two components were separated not in band space but in real space as in a Josephson junction, $\kappa_1$ and $\kappa_2$ would be the actual GL parameters of the two superconductors in contact. Note that $\bar{\kappa}$ varies with temperature contrary to $\kappa_n$. To discuss the different types of vortex interaction it is convenient to rescale the distance $x =\lambda\sqrt{2} x' $, the vector potential ${\bf A}= (\Phi_0/\sqrt{2} \pi \lambda ) {\bf a}$ and $\Psi_n = u_n \psi_n$. Here the magnetic penetration depth is defined by 
\begin{equation}
\lambda^2= \Phi_0 \bar{\kappa}/(4\pi)^{3/2}|\epsilon_M|^{1/2}.
\end{equation}
The normalized functional ${\cal F}\equiv ({\cal F}_{GL} - \epsilon_M)/\epsilon_{\lambda}$ is then 
\begin{eqnarray}
 {\cal F} & = & \frac{1}{2\pi} \{ \bar{\kappa}^2 \left[   p_1 (|\psi_1|^2\! - \! 1)^2  + p_2 (|\psi_2|^2 \!-\! 1)^2 + q |\psi_1 \! - \! \psi_2|^2 \right] \nonumber \\
 & + & r_1 |D\psi_1|^2 + r_2|D \psi_2|^2 + \frac{1}{2}(\nabla \times {\bf a})^2 \} 
\label{eq:norm_F} 
\end{eqnarray}
where the line energy $\epsilon_{\lambda}= (\Phi_0/4\pi\lambda)^2$, $D= (-i\nabla + {\bf a})$, the normalized Josephson coupling factor
\begin{equation}
q = \gamma u_1 u_2/|\epsilon_M|
\end{equation}
and the superfluid-density fraction 
\begin{equation} 
r_n= u_n^2 m_n^{-1} /(u_1^2 m_1^{-1}  +  u_2^2 m_2^{-1}).
\end{equation}
So in contrast to single-component superconductivity where the shape of the vortex interaction depends only on $\kappa$ \cite{jacobs}, the shape depends here on  four parameters: $\{ \bar{\kappa},p_1,q,r_1\}$ (since $p_1+p_2=1$ and $r_1+r_2=1$) or equivalently $\{ \bar{\kappa},\kappa_1,\kappa_2,q\}$. However the Josephson term locks $\psi_1$ with $\psi_2$ when $q$ diverges to infinity at the critical temperature (since $u_n$ vanishes) and, as illustrated below, the single-component behavior with $\psi_1=\psi_2$ and $\kappa=\bar{\kappa}$ is recovered in this limit if $\bar{\kappa} \ne 1/\sqrt{2}$.
  
\begin{figure}
%\begin{center}
 \scalebox{.72}{%
 \includegraphics*{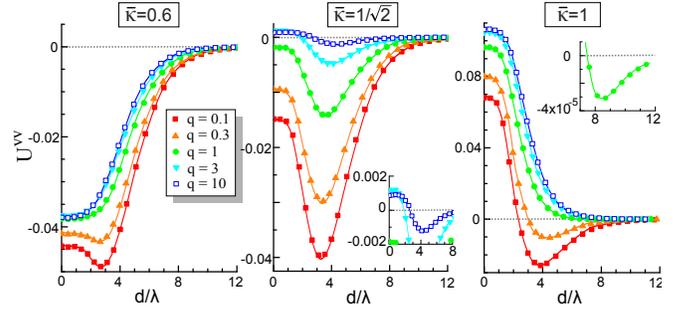}}
%\end{center}
\caption{(color online) Potential of the vortex interaction $U^{vv}(d)\equiv F(d)-F(\infty)$ as a function of vortex separation $d$  for different values of average GL parameter $\bar{\kappa}$ (band parameters $\kappa_1=0.5$, $\kappa_2=5$) and normalized Josephson coupling $q$.}
\label{fig:pot}
\end{figure}  
  
\emph{Numerical evaluation of the vortex interaction.\textemdash} The vortex-pair potential is numerically calculated by variational minimization of the GL energy $F=\int \! {\cal F} d^2\!x$ with the constraint that the pair separation $d$ is held fixed \cite{num_pot}. The \textit{Ansatz} \cite{jacobs} for $F$ is built upon isolated-vortex solutions with winding number $L=1$ and 2. For the latter the radial distributions $\psi_n(r)$ and ${\bf a}(r)$ are found by solving the finite difference version of the GL equations on a linear grid of 3000 points with a relaxation method. The boundary conditions are $|\psi_1|=|\psi_2|=a_{\theta}=0$ at $r=0$ and $|\psi_1|=|\psi_2|=1$, $a_{\theta}=-L/\sqrt{2}r$ at large distance $r\sim 100 \lambda$.

Fig.~\ref{fig:pot} shows the interaction potential $U^{vv}(d) \equiv F(d) - F(\infty)$ of two vortices when one OP component is of type-1 ($\kappa_1=0.5$) and the other one of type-2 ($\kappa_2=5$) while the average $\bar{\kappa}$ is 0.6, $1/\sqrt{2}$ and 1. For $\bar{\kappa}=0.6$ ($\bar{\kappa}=1$) the energy of a two-quanta vortex is smaller (larger) than the energy of two isolated vortices like in a single-component type-1 (-2) superconductor. But we note that this is not always the case for other choices of parameters. Besides for small $q$, the inter-vortex potential is repulsive at short separation and attractive at long range even though $\bar{\kappa}$ is not in the vicinity of $1/\sqrt{2}$. As $q$ increases the potential is first of type 1.5 and then it recovers a type-1 (-2) shape while the position $d_{\rm min}$ of its minimum goes to 0 ($+\infty$). In contrast, the potential for $\bar{\kappa} = 1/\sqrt{2}$ is always of type 1.5 whatever the strength of the coupling.     
 
\begin{figure}
%\begin{center}
 \scalebox{.58}{%
 \includegraphics*{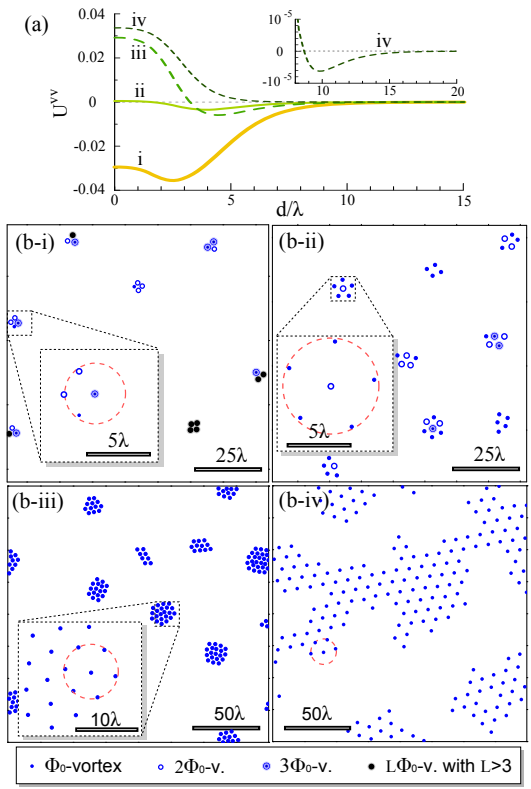}}
%\end{center}
\caption{(color online) (a) Different potential of vortex interaction in a type-1.5 superconductor. The inset is a magnification of curve {\it iv}. 
(b) The vortex distributions resulting from the interaction potentials {\it i} to {\it iv}. $L\Phi_0$-vortex means a vortex with $L$ quanta of flux. The dashed circle represents the positions of $U^{vv}$ minimum around one vortex.}
\label{fig:pattern1}
\end{figure}  

\emph{Distribution of vortices.\textemdash} To illustrate the variety of magnetic responses in a type-1.5 superconductor Fig.~\ref{fig:pattern1} shows the vortex patterns resulting from different interaction potentials $U^{vv}$ ranging from the more type-1-like ({\it i}) to the more type-2-like ({\it iv}). They are obtained by molecular dynamics  simulations of a system of over-damped vortices (see Ref.~\cite{moshchalkov} for details). Starting from several random distributions the dynamics of 200 vortices was simulated in a square of size $200\lambda \times 200 \lambda$ with periodic boundary conditions during 100000 time steps \cite{vort_md}.

\begin{figure}
%\begin{center}
 \scalebox{.64}{%
 \includegraphics*{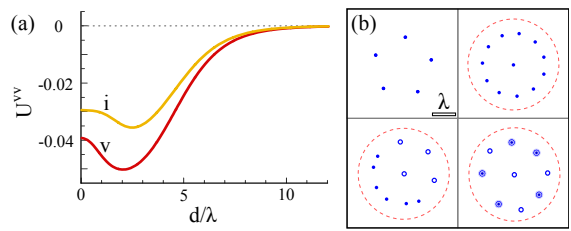}}
%\end{center}
\caption{(color online) (a) Vortex-interaction potentials for $\{ \bar{\kappa},\kappa_1,\kappa_2,q \} = \{0.65,0.5,25,0.4 \}$ (curve {\it i}) and   $\{0.83,0.5,100,0.08 \}$ (curve {\it v}). (b) Different vortex rings obtained with potential {\it v} and resulting from local variations of the vortex density (same conventions as in Fig.~\ref{fig:pattern1}).}
\label{fig:pattern2}
\end{figure}

The highly inhomogeneous vortex distributions are characteristically composed of areas filled with vortices surrounded by or surrounding vortex-free regions, depending on the ratio of the minimal-potential separation $d_{\rm min}$ to the average distance $d_{vv}$ between vortices (compare e.g. (b-iii) with (b-iv) in Fig.~\ref{fig:pattern1}). When $d_{\rm min} > d_{vv}$ the whole surface is covered by a vortex lattice even though the interaction is of type-1.5. With potentials {\it i} and {\it ii} a multi-quantum vortex can be stabilized by the surrounding vortices at distances $\sim d_{\rm min}$: the latter generate a local potential well which overcomes the repulsion between the merged vortices forming the giant vortex. However the maximum of $U^{vv}$ at zero separation prevents all the vortices from collapsing into one unique macroscopic vortex, i.e. a normal region like in the intermediate state. Giant vortices can be observed in mesoscopic samples \cite{kanda} or trapped by pinning sites \cite{buzdin}. But the ability to stabilize them in the bulk without any extrinsic confinement mechanism is one unique property of type-1.5 superconductivity. Yet we found an even more striking feature: rings of vortices can appear, as shown in Fig.~\ref{fig:pattern2}, when the OP components vary on length scales which are greatly different (i.e. when $\kappa_1 \ll \kappa_2$ and $q \ll 1$). In this situation the local maximum of the inter-vortex potential at zero separation is sharper than the minimum (see curves {\it i} and {\it v} in Fig.~\ref{fig:pattern1} and \ref{fig:pattern2}) which is detrimental to the stability of a giant vortex. The favored configuration is then an annular distribution of radius $< d_{\rm min}$, with the presences of one central vortex and of multi-quantum vortices if the vortex density is high.

\emph{Limit of large Josephson coupling.\textemdash} The robustness of type-1.5 superconductivity at $\bar{\kappa}\approx 1/\sqrt{2}$ towards the large coupling $q$ can be analytically demonstrated.  When $q$ is large the difference $(\psi_1 -\psi_2)$ is small so the GL functional (\ref{eq:norm_F}) can be minimized with a perturbative expansion in powers of $q^{-1}$ by writing $\psi_{1,2}= \varphi_0 \pm \varphi_1 /2q$. At zeroth order in $q^{-1}$, $\varphi_0$ and $a$ are then solutions of the Bogomolnyi-Jacobs-Rebbi (BJR) equations when $\bar{\kappa}\approx 1/\sqrt{2}$ \cite{bogomolnyi,jacobs,lukyanchuk} while the first-order correction $\varphi_1= (P - R/2\bar{\kappa}^2)(1-|\varphi_0|^2) \varphi_0$ with 
\begin{equation}
P= p_1 - p_2 \;\; {\rm and} \;\; R= r_1 - r_2.
\end{equation}        
At zeroth order the free energy does not vary with the distance between the vortices since the energy of BJR solutions does not depend on the positions of the vortices. Its dependence on the separation is given at first order of correction by the variation of $\int \!\! d^2\!x [ (\bar{\kappa}^2-1/2 + 2 s) |\varphi_0|^4  - s  |\varphi_0|^6 ]$ which can be estimated in the limits of short and large separation as in Ref.~\cite{lukyanchuk}. Here $s=(2 \bar{\kappa}^2 P - R)^2/4 q \bar{\kappa}^2$. We find \cite{bounds} that for large Josephson coupling $q$ the vortex interaction is 
\begin{eqnarray}
{\rm of \; type \; 2 \;  when \;} & C\equiv\frac{ q  (1 - 1/2 \bar{\kappa}^2)}{2(P - R/2 \bar{\kappa}^2)^2} > 0.346, \nonumber \\
{\rm of \; type \; 1.5 \;   when \;} & -0.4 < C < 0.346, \nonumber \\
{\rm of \; type \; 1 \;  when \;} & C <-0.4. 
\label{eq:C}
\end{eqnarray}
$P$ and $R$ quantify the disparities between the two bands. When they are small or when $q$ is large, as in the vicinity of $T_c$, the ratio $C$ grows out of the bounds defining type~1.5 and superconductivity is either of type~1 or of type~2. However in the case $\bar{\kappa}=1/\sqrt{2}$, $C$ is exactly zero which means that the pair potential is always non-monotonic independent of the strength of the Josephson coupling.

\begin{figure}
%\begin{center}
 \scalebox{.44}{%
 \includegraphics*{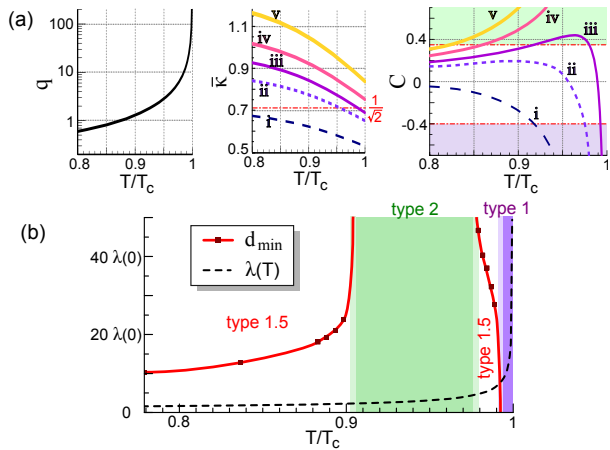}}
%\end{center}
\caption{(color online) (a) Temperature evolution of normalized Josephson coupling $q$, average GL parameter $\bar{\kappa}$ and vortex-interaction parameter $C$ for $\{ \kappa_1,\kappa_2 \}=$ (i) $\{ 0.5,1.5 \}$, (ii) $\{ 0.6,2 \}$, (iii) $\{ 0.6,2.65 \}$, (iv) $\{ 0.65,3 \}$  and (v) $\{ 0.7,4 \}$. (b)~Temperature evolution of the separation $d_{\rm min}$ at $U^{vv}$-minimum and of the magnetic penetration depth $\lambda$ for the parameter set (iii) in (a).}
\label{fig:temp_evo}
\end{figure}

\emph{Temperature dependence of the interaction.\textemdash} Since the average $\bar{\kappa}$ and the normalized Josephson coupling $q$ depend on temperature the vortex-vortex interaction can also change with it. As an illustration,  the temperature evolution of these quantities and of the ratio $C$ are plotted in Fig.~\ref{fig:temp_evo}(a) for different values of $\{ \kappa_1,\kappa_2 \}$ which yield type-1.5 superconductivity away from $T_c$. We used parameters which are in the range of experimental estimates for MgB$_2$ \cite{moshchalkov,eisterer,parameters}. We adopted the model of temperature dependence for $\alpha_n$ from Ref.~\cite{eisterer,dao} with $\alpha_1$ becoming negative below $T_1=0.64 T_c$ and $\alpha_2$ below $T_2=0.91 T_c$. At the critical temperature $q$ diverges to infinity so the vortex-vortex interaction is either of type~1 or type~2 just below $T_c$. For the majority of parameter sets there is only one change of behavior type-1(-2)$\rightarrow$type-1.5 as temperature decreases. But as shown by the example (iii) in Fig.~\ref{fig:temp_evo}(a) $\bar{\kappa}$ can pass by the value $1/\sqrt{2}$ in the vicinity of $T_c$ while $q$ is large. In this case $C$ can go out of the bounds~(\ref{eq:C}) twice which means an alternation of four regimes is possible. It was previously found \cite{moshchalkov,babaev2} that a type-1.5 phase can survive above min$(T_1,T_2)$ but a second interval above max$(T_1,T_2)$ was unexpected. This four-phase scenario is deduced from inequalities~(\ref{eq:C}) that are valid only when $q$ is large. Thus $U^{vv}$ for the set of parameters (iii) is numerically evaluated to check the estimation. We find that the phase diagram is indeed composed by an unusual reentrant succession of four regimes: type-1$\rightarrow$type-1.5$\rightarrow$type-2$\rightarrow$type-1.5. The temperature dependence of the position $d_{\rm min}$ and of the penetration depth $\lambda$ are plotted in Fig.~\ref{fig:temp_evo}(b). When the superconductor enters a type-1 phase the position of the non-monotonic-potential minimum goes to zero while it diverges to infinity when entering a type-2 phase.

In conclusion, we predict novel properties for type-1.5 superconductivity which strikingly distinguish it from the vortex states of conventional type-1 and type-2 superconductors. By numerical GL calculations of the vortex interaction potential and simulations of the resulting vortex dynamics, we found vortex-stabilized giant vortices which survive in bunches and unusual vortex rings in bulk multiband superconductors. The  multi-component character produces as well a complex phase diagram where type-1.5 behavior can re-enter by alternating with type-1 and type-2 interludes in a succession of three regime transitions as temperature is changed.

VHD acknowledges the financial support by the grant EF/05/005 (INPAC) from the 
University of Leuven.


\begin{thebibliography}{30}
\bibitem{num_pot} The conjugate gradient method is stopped when the relative difference of energy between successive iterations is  $<5\times 10^{-8}$. When compared to the exact value $F(d)=2$ for a single component at $\kappa=1/\sqrt{2}$ \cite{jacobs} the relative error is $\approx 10^{-5},10^{-4}, 10^{-5}$ and $10^{-7}$ at $d/\lambda= 1,2,5$ and 10.
\bibitem{vort_md} The local ordering inside a vortex cluster is extremely fast in comparison to its migration because the inter-vortex force decreases exponentially with the separation. The stability of the final configuration was tested by running the simulation from this configuration for 200000 more time steps with step duration 4 times larger.
\bibitem{bounds} The upper bound is equal to $(1 - \delta \!\! \int\! |\nabla \! \times \! a|^3/\delta \!\!\int\! |\nabla \!\times\! a|^2)/2$ that is numerically estimated to $0.346$ ($\delta \!\! \int$ stands for the variation of the integral as $d$ grows to infinity). The lower bound $3 (\zeta_8^{(2)}-\zeta_6^{(2)})/4(\zeta_6^{(2)}-\zeta_4^{(2)}) - 1 \approx -0.4$ with the values of the constants $\zeta_n^{(2)}$ given by Ref.~\cite{lukyanchuk}.
\end{thebibliography}
\end{document}